# Photothermally induced, reversible phase transition in methylammonium lead triiodide

*Shunran Li[1,2], Zhenghong Dai[3], Conrad A. Kocoj[1,2], Eric I. Altman[1], Nitin P. Padture[3], Peijun Guo[1,2,*]*

[1]Department of Chemical and Environmental Engineering, Yale University, 9 Hillhouse Avenue, New Haven, CT 06520, USA

[2]Energy Sciences Institute, Yale University, 810 West Campus Drive, West Haven, CT 06516, USA

[3]School of Engineering, Brown University, Providence, RI 02912, USA

*Contact author: peijun.guo@yale.edu

**Abstract**

Metal halide perovskites (MHPs) are known to undergo several structural phase transitions, from lower to higher symmetry, upon heating. While structural phase transitions have been investigated by a wide range of optical, thermal and electrical methods, most measurements are quasi-static and hence do not provide direct information regarding the fundamental timescale of phase transitions in this emerging class of semiconductors. Here we investigate the timescale of the orthorhombic-to-tetragonal phase transition in the prototypical metal halide perovskite, methylammonium lead triiodide ($CH_3NH_3PbI_3$ or $MAPbI_3$) using cryogenic nanosecond transient absorption spectroscopy. By using mid-infrared pump pulses to impulsively heat up the material at slightly below the phase-transition temperature and probing the transient optical response as a function of delay time, we observed a clean signature of a transient, reversible orthorhombic-to-tetragonal phase transition. The forward phase transition is found to proceed at tens of nanoseconds timescale, after which a backward phase transition progresses at a timescale commensurate with heat dissipation from the film to the underlying substrate. A high degree of transient phase transition is observed accounting for one third of the steady-state phase transition. In comparison to fully inorganic phase-change materials such as $VO_2$, the orders of magnitude slower phase transition in $MAPbI_3$ can be attributed to the large energy barrier associated with the strong hydrogen bonding between the organic cation and the inorganic framework. Our approach paves the way for unraveling phase transition dynamics in MHPs and other hybrid semiconducting materials.

Metal-halide perovskites (MHPs) have emerged in the past decade as a promising class of semiconducting materials for photovoltaic and optoelectronic applications.[1-5] Their appealing features include facile solution processability, low material cost, and defect tolerance to enable exceptional device performance. The possibility of incorporating organic spacers in MHPs to form reduced-dimensional structures has lead to remarkable structural diversity, and with it a wide range of properties not only for photovoltaics and optoelectronics,[6] but also for lateral heterojunctions, spintronics and optical chirality.[3, 7, 8] Unlike conventional semiconductors such as Si and GaAs, MHPs, similar to oxide perovskites, undergo a sequence of phase transitions, typically going from a less symmetric space group at low temperatures to a higher symmetry space group upon heating.[9] Understanding the phase transitions in MHPs is crucially important since phase transitions are naturally encountered in the solution synthesis and processing of these materials, as well as their degradation and failure over time.[10] On the other hand, manipulating phase transitions of MHPs offers a unique 'tuning knob' of the material properties such as optical transparency and stimuli-responsiveness.[11-15]

Phase transitions in MHPs have been investigated in the past by a myriad of techniques such as X-ray diffraction,[16] time-of-flight neutron scattering,[17] Raman spectroscopy,[18] and differential scanning calorimetry.[19] Recently, the kinetics and energetics of phase transitions between the non-perovskite and perovskite phases for several MHPs have been explored using advanced micro-photoluminescence spectroscopy and cathodoluminescence techniques,[20-22] and it was revealed that a liquid-like interface between the two disparate phases facilitates the phase transition owing to configurational entropy. Nevertheless, most investigations on phase transitions of MHPs have temporal resolutions in the second range, and fundamental timescales of phase transitions in prototypical MHPs, such as methylammonium lead triiodide ($CH_3NH_3PbI_3$, or $MAPbI_3$), remain unexplored so far. Developing new experimental schemes with high temporal resolution can help unravel the fundamental timescales of phase transition in this technologically important class of materials.

In this work, we report on an ultrafast, optical pump-probe spectroscopic study of a phase transition in $MAPbI_3$. We leverage the strong vibrational absorption of this material in the mid-infrared range (near 3200 cm$^{-1}$), which permits strong vibrational excitation to transiently heat up the material in an impulsive fashion. Following the mid-infrared pump with an electronically-delayed broadband probe pulse, we observe a clean and unambiguous signature, attributable to the orthorhombic-to-tetragonal phase transition. This phase transition is found to proceed in the tens-of-ns timescale, which is significantly slower than phase transitions in inorganic materials such as correlated oxides.[23] Comparing transient spectra with steady-state data reveals that in a spatially averaged sense, the transiently induced phase transition can account for as much as one third of a full steady-state phase transition, which is currently limited by the competing heat transport process from the film to the substrate. Our method can be generalized to the study of other hybrid semiconductors exhibiting phase transitions and opens up pathways for realizing optically triggered and controlled switching functionalities when the two distinct phases have disparate electronic, optical, or magnetic properties.

**Steady-state characterization**

Figure 1a presents the steady-state optical transmittance measured for a ~600-nm thick MAPbI$_3$ sample as a function of temperature. The MAPbI$_3$ film is deposited on a mid-infrared transparent CaF$_2$ substrate, with a cross-sectional scanning electron microscopy (SEM) image shown in Fig. 1b. As the temperature increases from 78 K, the MAPbI$_3$ film is initially in the orthorhombic phase with a space group of *Pnma*.[16, 17, 24] At about 158 K, a structural phase transition to the tetragonal phase with an I4/*mcm* space group is observed, which is associated with an abrupt redshift of the absorption onset wavelength from ~745 nm to ~790 nm as seen in the transmittance data.[25] The crystal structures of the two relevant phases of MAPbI$_3$ in this work are sketched in Fig. 1c. The first-order, orthorhombic-to-tetragonal phase transition involves the rotation of Pb-I octahedra (a combination of in-plane rotation along the *b* axis and out-of-plane rotations along the *a* and *c* axes), coupled with an order-disorder transition of the MA cations.[17, 26] The MA cations with C$_{3v}$ point group are frozen in the orthorhombic phase, but undergo liquid-like reorientational motions in the tetragonal phase with characteristic timescales of a few ps.[27, 28] In both the orthorhombic and tetragonal phases, the material's optical absorption onset wavelength blueshifts with increasing temperature.

The signature of the phase transition is also reflected in steady-state, temperature dependent infrared absorption spectra shown in Fig. 1d. Specifically, when transitioning from orthorhombic to tetragonal, a strong absorption near 3100 cm$^{-1}$ arising from the asymmetric N-H stretching vibrations drops to nearly zero, although some vibrational absorption near 3200 cm$^{-1}$ from the $CH_3NH_3^+$ cations remains active in the tetragonal phase.[29] Such vibrational absorption of $MAPbI_3$ in the mid-infrared (~0.4 eV) furnishes us a possibility of pure vibrational (*i.e.*, photothermal) stimulation of the material in optical pump-probe spectroscopy measurements (Fig. 1d, inset). Comparing to the above-bandgap pumping scheme widely adopted in pump-probe measurements, mid-infrared pumping does not involve the excitation of charge carriers from the valence to the conduction band, and hence can provide a clean signature of photoinduced transient phase transitions, as reported in this work.

In our ns-μs transient absorption measurements, the samples were photothermally excited by mid-infrared pump pulses centered at 3170 nm with a pulse width of ~170 fs and a spectral full-width-half-maximum of about 200 $cm^{-1}$. Broadband probe pulses covering a spectral range from 450 nm to 900 nm have a pulse width of 1~2 ns. The overall instrument response function of our setup is about 4 ns (see Supplementary Fig. S1). As we have shown in past work focusing on fs-to-ns transient absorption,[30] mid-infrared pump pulses at ~3170 nm resonantly and transiently populate the N-H antisymmetric stretching vibrations of the organic $CH_3NH_3^+$ cations, which subsequently relax by dissipating the excess vibrational energy into the inorganic Pb-I octahedral framework. Following fs mid-infrared pulsed laser excitation, the organic and inorganic sublattices reach a mutual thermal equilibrium via phonon-phonon interactions in a sub-ns timescale. As such, during the ns-to-μs time window investigated in this work, the various phonon modes of $MAPbI_3$, whose frequencies span from 3200 $cm^{-1}$ on the organic cations down to nearly zero wavenumber for the inorganic sublattice,[18] maintain an internal thermal equilibrium. The relevant processes here are heat-induced phase transition, concurrent with heat transfer from $MAPbI_3$ to the substrate, since after the vibrational excitation the $MAPbI_3$ film occupies a higher temperature than the $CaF_2$ substrate. Note that the $CaF_2$ substrate is transparent to mid-infrared excitation and hence remains cold relative to the thermally excited $MAPbI_3$ film before heat transfer takes place.

**Transient results far from the phase transition**

Figure 2 presents the measured transient $\Delta T/T$ spectral map for the $MAPbI_3$ film on $CaF_2$ at several selected temperatures. Here $\Delta T/T$ is defined as $\frac{T(t)-T(0)}{T(0)}$, where $T(0)$ is the transmittance without the pump excitation, and $T(t)$ is the transmittance at delay time $t$ following the pump excitation. At 78 K (Fig. 2a), far from the orthorhombic-to-tetragonal phase transition temperature (denoted as $T_{o-t}$ hereafter), we see an asymmetric, derivative-like spectral feature which undergoes significant decay in a few-hundred-ns time window. Specifically, photothermal excitation leads to enhanced transmission on the red side and reduced transmission on the blue side with respect to the $\Delta T/T$ zero-crossing wavelength, which is centered at the steady-state absorption peak wavelength of 740 nm. Comparing such transient spectral feature with the steady-state transmittance shown in Fig. 1a, we find the observed $\Delta T/T$ response corresponds to a transient lattice temperature rise: the increase in lattice temperature causes a blueshift of the absorption onset as well as a more diminished exciton absorption peak intensity in the orthorhombic phase (see Supplementary Fig. S2 for a plot in terms of the optical density, which offers a higher contrast near the exciton absorption peak). The increase in lattice temperature is maximal at earliest delay time (i.e., at about 1 ns when the organic and inorganic sublattices just reach mutual thermal equilibrium before macroscopic transfer) and gradually decays over time due to heat transfer from $MAPbI_3$ to the substrate. Using reported values for the thermal conductivities and heat capacities of $MAPbI_3$ and $CaF_2$ at 80 K, we simulated the characteristic timescale for heat transfer from an initially uniformly heated $MAPbI_3$ to an unheated $CaF_2$ substrate (Supplementary Fig. S3). The simulated decay of the average temperature of $MAPbI_3$ film agrees well with experimentally observed few-hundred-ns timescale of the $\Delta T/T$ signal.

Fig. 2b shows the transient result measured at 162 K in the tetragonal phase (162 K is above $T_{o-t}$). The positive $\Delta T/T$ signal corresponds to an increased transmittance, which, once compared with the steady-state transmittance data in Fig. 1a, again indicates an impulsive lattice temperature rise followed by a decay. Similar to the data for 78 K (Fig. 2a), the elevated lattice temperature relaxes in a few hundred ns. We note that the transient $\Delta T/T$ spectral map measured at 162 K is different in shape from its counterpart at 78 K, as no derivative-like line shape is seen for the spectra at 162 K. We ascribe such difference to the following: although blueshifts in the absorption onset wavelengths are observed for both phases, there is a lack of an apparent exciton absorption resonance in the tetragonal phase, and as a result, a blueshift in the absorption onset should only give rise to a transiently increased transmittance. Equivalently speaking, a derivative-like $\Delta T/T$ line shape must come from the shift of a resonant steady-state feature, which only holds true for the orthorhombic phase due to its strong excitonic absorption.

**Transient results near the phase transition**

After examination of photothermally induced transient behavior of MAPbI$_3$ away from the phase transition, we then performed measurements at temperatures slightly below $T_{o-t}$. We hypothesized that if the lattice temperature transiently climbs above the phase transition point, the phase transition should be inducible. Fig. 2c displays the $\Delta T/T$ spectral map measured at 142 K, and the result drastically differs from those shown in Fig. 2a-b. Notably, the spectral feature is composed of a positive narrow $\Delta T/T$ signal centered at 735 nm, as well as a much broader, negative $\Delta T/T$ signal centered at about 760 nm. The positive $\Delta T/T$ feature has a fast rise in the first few ns (falling below the instrument response of our setup), which is similar to the fast rise time of $\Delta T/T$ observed at both 78 K and 162 K. In contrast, the time evolution of the negative $\Delta T/T$ signal has a much longer rise time in the ~100 ns regime, suggesting that a different photoinduced process is at play. At 142 K, MAPbI$_3$ is in the orthorhombic regime (Fig. 1a), hence upon photothermal excitation an impulsive temperature rise, followed by a temperature decay, can account for the positive $\Delta T/T$ band centered at 735 nm. As expected, the decay timescale of the positive $\Delta T/T$ signal matches with the results obtained at 78 K and 162 K. In order to determine the origin of the broadband negative $\Delta T/T$ feature, we can revisit the steady-state, temperature-dependent transmittance data in Fig. 1a. Notably, accompanying the steady-state orthorhombic-to-tetragonal phase transition is an abrupt redshift of the absorption onset wavelength (from ~740 nm to ~780 nm), so a photoinduced phase transition is expected to lead to an enhanced absorption in the range of 740 nm ~ 780 nm, which matches very well with the negative $\Delta T/T$ signal in Fig. 2c. As a result, we can attribute the broad negative $\Delta T/T$ band to a photothermally induced orthorhombic-to-tetragonal phase transition. Since 142 K is still relatively far from $T_{o-t}$, the most negative $\Delta T/T$ value induced by photothermal excitation reaches about -1.5%, indicating that only a small fraction of the material undergoes the induced phase transition. We note that since our transient absorption experiments were performed at 1500 Hz (750 Hz) laser repetition rate for the probe (pump) and the measurement of each $\Delta T/T$ transient spectral map takes 10 to 15 minutes; the observed transient phase transition is fully reversible.

To gain more insights into the photothermal induced phase transition, we performed transient absorption experiments at every two Kelvin in the range of 142 K to 160 K (the full transient $\Delta T/T$ spectral maps are shown in Supplementary Fig. S4 and S5), above which the material completely transitions into tetragonal. Fig. 2d presents the data acquired at 154 K, which is very close to $T_{o-t}$.

The result in Fig. 2d is qualitatively similar to that obtained at 142 K, in that the $\Delta T/T$ is mainly composed of a positive band centered at 728 nm and a negative band centered at 757 nm. In addition, we observe a minor, third feature, which is a short-lived positive $\Delta T/T$ signal spanning from 760 nm to 800 nm. Because this third feature has a fast rise time that is similar to the positive $\Delta T/T$ signal arising from heat dissipation of the orthorhombic phase, we can attribute it to transient impulsive lattice temperature rise and subsequent temperature dissipation of the pre-existing minor tetragonal phase in the matrix. Note that at 154 K, the steady-state transmittance data (Fig. 1a) indicates that some minor tetragonal phase has already formed in the orthorhombic matrix. The much faster decay time of the third $\Delta T/T$ feature, however, arises since this positive $\Delta T/T$ signal spectrally overlaps with, and is largely overwhelmed by, the broadband, phase-transition induced negative $\Delta T/T$ signal centered at 757 nm with a much stronger amplitude. Note that at 156 K and 158 K (Supplementary Fig. S6), this positive $\Delta T/T$ band from 760 nm to 800 nm exhibits a longer decay time and larger amplitude, since a growth in the equilibrium fraction of tetragonal phase leads to a larger $\Delta T/T$ amplitude and associated with it a diminishing degree of photothermally induced orthorhombic-to-tetragonal phase transition.

Comparing the positive $\Delta T/T$ signal centered at 728 nm measured at 154 K (Fig. 2d) and its counterpart centered at 735 nm obtained at 142 K (Fig. 2c), we find that the former signal undergoes much smaller decay over the plotted 400 ns time window. The $\Delta T/T$ kinetics at 757 nm represents transient, photothermally induced forward and backward phase transitions, whereas the $\Delta T/T$ kinetics at 728 nm depicts temporal evolution of the fraction of the orthorhombic phase that has not undergone any phase transition. At 154 K, the photothermally induced phase transition leads to a maximal transmittance loss of 8% at 757 nm, which is substantially higher than that seen at 142 K (~1.5%). As a result, the tetragonal-to-orthorhombic backward phase transition, which kicks in at around 200 ns after the signal from transiently formed tetragonal phase reaches its plateau, "replenishes" the higher-temperature (with respect to the lattice temperature before excitation) orthorhombic phase, thereby contributing to a positive $\Delta T/T$ signal centered at 728 nm and leading to a slowdown in its decay in comparison to the result collected at temperatures further away from the phase transition (e.g., Fig. 2a-c).

**Temperature dependent kinetics**

To further examine the amplitude and kinetics of $\Delta T/T$ associated with the transient phase transition, we plot the temperature dependent decay kinetics of $\Delta T/T$ in Fig. 3a and 3b, extracted at the wavelengths of 757 nm and 728 nm, respectively. From Fig. 3a we see that the photothermally induced phase transition is clearly observed at temperature as low as 130 K, which is about 26 K below the steady-state thermodynamic $T_{o-t}$ (156~158 K). It is seen that as the temperature approaches $T_{o-t}$ from below, the strength of the negative $\Delta T/T$ signal grows in a superlinear fashion with temperature. Although a constant pump fluence of 2.4 mJ·cm$^{-2}$ was used in these temperature dependent measurements, the absorbed pump fluence, and the resulting transient rise of lattice temperature near time zero, is expected to decrease with increasing measurement temperature based on the steady-state infrared absorption spectra which show a decreasing absorption of the mid-infrared pump (Fig. 1b). The superlinear growth of the transiently formed tetragonal phase reflects a similar picture in the steady-state condition, that is, a superlinear growth of the volume fraction of the incipient tetragonal phase in the orthorhombic matrix with linearly increasing temperature. The presence of ionic defects[31, 32] and grain and twin boundaries[33] typically found in MHPs can facilitate the formation of tetragonal phase below the $T_{o-t}$. Spatially resolved X-ray diffraction experiments,[34] either static or time-resolved, can provide more direct and conclusive evidence regarding the spatial distribution of the activation energy lowering by the incipient tetragonal phase domains in the orthorhombic matrix. When the measurement temperature reaches 156 K, a decline in the amplitude of the negative $\Delta T/T$ is observed, which corresponds to a diminishing steady-state orthorhombic volumetric fraction in favor of nearly complete formation of the tetragonal phase.

The temperature dependent $\Delta T/T$ kinetics at 728 nm in Fig. 3b capture the temporal evolution of the orthorhombic phase. At low temperatures (130 K to 140 K), the fast rise in $\Delta T/T$ (<5 ns) is followed by a nearly mono-exponential decay in the time window up to 600 ns, reflecting heat transfer. As temperature increases from 144 K, the fast $\Delta T/T$ rise starts to be followed by a much slower rise, the timescale of which quantitatively matches with the growth of the negative $\Delta T/T$ signal in Fig. 3a at the corresponding temperatures. The agreement between the two timescales (i.e., slow growth of the negative $\Delta T/T$ signal at 757 nm and slow rise of the positive $\Delta T/T$ at 728 nm) stems from the fact that the transmittance of MAPbI$_3$ at 728 nm is higher in the tetragonal phase than in the orthorhombic phase (see Fig. 1a and Supplementary Fig. S2) due to the stronger excitonic absorption by the latter, so a transient orthorhombic-to-tetragonal phase transition leads to a slowly increased transmittance at 728 nm. At longer delay time, a slower $\Delta T/T$ decay is seen at higher temperature, due to the backward phase transition as discussed before. We note that the backward phase transition generally proceeds at a slower pace in comparison to the forward phase transition, since even at 600 ns the effects from backward phase transition can still be seen especially at higher measurement temperatures. The slower backward phase transition stems from the inefficient heat dissipation from the MAPbI$_3$ film to the substrate due to the low thermal conductivity of MAPbI$_3$,[35] which is slower than the forward phase transition timescale.

Figure 4a shows fluence-dependent transient $\Delta T/T$ spectra obtained at ~200 ns, when the signal at 757 nm reaches the most negative values corresponding to the maximal degree of the forward phase transition. The $\Delta T/T$ signal at 757 nm exhibits a superlinear dependence on pump fluence, which essentially has an analogous origin to the superlinear dependence of $\Delta T/T$ at 757 nm with the measurement temperature (Fig. 3a). Specifically, it arises since the tetragonal fraction in MAPbI$_3$ grows superlinearly with lattice temperature near the phase transition, and the lattice temperature rise scales linearly with the pump fluence. As expected, the fluence dependent $\Delta T/T$ kinetics at 757 nm (in Fig. 4b) shows a longer rise time for $\Delta T/T$ to reach the negative peak. By comparing the transient $\Delta T/T$ spectra at the delay time when the negative $\Delta T/T$ band reaches its peak value (i.e., at ~200 ns) with the $\Delta T/T$ spectra calculated using the steady-state transmittance data with respect to the transmittance at 154 K (Supplementary Fig. S6), we can estimate the relative degree of phase transition in the system. With a complete phase transition, the amplitude of $\Delta T/T$ at 757 nm calculated from the steady-state data (Supplementary Fig. S8) reaches about -0.6, whereas the $\Delta T/T$ from transient measurements reaches about -0.2 at the highest fluence of 5.6 mJ·cm$^{-2}$, indicating that about one third of the film has undergone the transient phase transition. We note the tendency toward further transient phase transition beyond this ratio is suppressed by the inevitable film-to-substrate heat transfer which leads to a reduction in the lattice temperature rise. Considering the relatively high thermal conductivity of CaF$_2$ (>10 W·m$^{-1}$·K$^{-1}$), using single crystals of MAPbI$_3$ or deposition of an MAPbI$_3$ film on very low thermal conductivity substrates such as polyethylene terephthalate (<0.3 W·m$^{-1}$·K$^{-1}$)[35] may improve heat retention for the development of a more complete transient phase transition.

To examine if the observed timescale of photothermally induced phase transition of MAPbI$_3$ grown on CaF$_2$ using the antisolvent method is a generic behavior, we performed additional measurements on an MAPbI$_3$ film made by the antisolvent-bath method on a *c*-plane sapphire substrate (see Methods). Steady-state transmittance data (Supplementary Fig. S7) shows that the $T_{o-t}$ of this film is about 134 K. Upon cooling-heating cycles, this sample shows a stronger hysteresis behavior than the film on CaF$_2$, and the $T_{o-t}$ slightly varies by a couple of degrees over the course of several heating-cooling cycles. The $\Delta T/T$ kinetics at 757 nm are presented in Fig. 3c and $\Delta T/T$ kinetics at 728 nm in Fig. 3d (full transient $\Delta T/T$ spectral maps are shown in Supplementary Fig. S8). Based on mono-exponential fits to the negative $\Delta T/T$ signal growth, we find this sample exhibits faster rate of phase transition (within the same order of magnitude), but the degree of transient phase transition is notably lower than MAPbI$_3$ on CaF$_2$, which is further evident from the drastically different $\Delta T/T$ response at 728 nm shown in Fig. 3d in comparison to Fig. 3b. The overall weaker $\Delta T/T$ amplitude observed for MAPbI$_3$ on sapphire can be largely attributed to the higher thermal conductivity of sapphire (about 4~5 times) in comparison to CaF$_2$ which more rapidly suppresses the forward transient phase transition. Future work is warranted to unravel the correlation between the transient phase transition kinetics and defects, strain, and surface-to-volume ratio of the film, which were found to influence $T_{o-t}$.[36-38]

**Transient response with a visible pump**

Considering that carrier thermalization in MAPbI$_3$ *via* carrier-phonon interactions and carrier-impurity scattering,[39-43] which are relevant processes in device operations, can lead to lattice heating, we further explored whether the transient phase transition can result from carrier thermalization by employing above-bandgap photoexcitation at 700 nm in transient absorption measurements. The $\Delta T/T$ spectral map acquired at 148 K is shown in Fig. 4c, and several representative decay kinetics are plotted in Fig. 4d. Firstly, a positive $\Delta T/T$ bleach signal is observed to center at 728 nm. This signal has an instantaneous rise, followed by an initial rapid decay that is notably faster than the counterpart measured under mid-infrared pump (Fig. 3b). We attribute this faster decay to rapid charge carrier recombination. In addition, we observe a broadband photoinduced absorption at wavelengths bluer than 700 nm above the bandgap (Fig. 4d, cyan curve). Such above-bandgap photoinduced absorption in tetragonal MAPbI$_3$ has been reported previously to arise from a transient reflectivity change of the film,[44] which should underpin the similar observation here for the orthorhombic phase. Most interestingly, at 757 nm, a rapid rise in $\Delta T/T$ is observed, which is followed by a decay down to its negative peak value at about 100 ns. We note that this ~100-ns timescale matches well with that measured under mid-infrared pumping (Fig. 3a), indicating that above-bandgap optical pumping also triggers a transient phase transition. The fact that a transient phase transition induced by above-bandgap pumping occurs at a similar rate as that driven by mid-infrared pumping suggests a common thermal origin behind these observations. Although the pump fluence in our transient measurements was in the mJ·cm$^{-2}$ regime, due to charge carrier localization in the form of excitons in orthorhombic MAPbI$_3$ and its small exciton Bohr radius (~5 nm),[45] we expect local photo(thermally) induced orthorhombic-to-tetragonal phase transition to be possible even at low levels of photoexcitation, which may impact the transient optoelectronic properties of charge carriers.[46] Relatedly, it was shown that photoinduced phase transitions in orthorhombic MAPbI$_3$ can generate local tetragonal inclusions to enable continuous-wave lasing.[47]

Photoinduced phase transitions in solids have been under intense investigation, as such studies can provide fundamental understanding of nonequilibrium states and hidden metastable phases of matter which are essential to achieve novel material properties for applications in optical switchable devices.[23, 48-50] When fs laser pulses are used as the driving force, photoinduced *adiabatic nonthermal* phase transitions typically occur in the ps timescales. Examples include the metal-to-insulator transition in the prototypical strongly correlated vanadium dioxide,[23, 51] phase transition and metastable phase formation in manganites,[50, 52] paraelectric-to-ferroelectric transition in SrTiO$_3$,[51, 53] and various chalcogenides[48, 54] and charge-density-wave systems.[55] Even for MHPs, a photoinduced orthorhombic-to-cubic phase transition was observed in all-inorganic perovskite CsPbBr$_3$ nanocrystals with ps transition timescales and a ~500 ps recovery time.[56]

The investigation of *photothermally* driven phase transitions, on the other hand, has been less reported. This is in part because of the lack of a suitable optical pumping scheme that can resonantly perturb the material without charge carrier excitation, as most inorganic materials have phonon vibrations at the far-infrared to terahertz regime where pulsed laser sources are much lower in peak intensity than the mid-infrared pulses used here. Our result reported here shows that the reversible orthorhombic-to-tetragonal phase transition in MAPbI$_3$ is at least three orders of magnitude slower than most of the previously reported photoinduced phase transitions in inorganic solids. The extremely slow transition can be largely attributed to the high energetic barriers against the cooperative orientational unlocking of the CH$_3$NH$_3^+$ cations[57] due to the strong hydrogen bonding between the organic cations and the iodide anions. Our report suggests that in other soft material systems with phase transitions that involve breaking of hydrogen bonding networks, the timescale is likely to be slow as well. The reported technique can be adapted for studying fundamental timescales and energetic barriers of phase transitions in several important MHP compositions beyond MAPbI$_3$, such as formamidinium lead triiodide (FAPbI$_3$), CsPbI$_3$, and CsSnI$_3$, to name a few.[58-60] For CsPbI$_3$ and CsSnI$_3$ that lack organic cations for mid-infrared resonant pumping, one strategy is to measure nanocrystals of these inorganic MHPs embedded in liquid or solid matrix made of organic materials. Information from future experiments of this kind can be useful for understanding and developing strategies to prevent undesirable phase transitions in MHPs for the fabrication of stable and efficient photovoltaic devices.

In conclusion, using vibrational-pump visible-probe spectroscopy in a ns-to-μs delay time window, we, for the first time, observed photothermally-induced transient orthorhombic-to-tetragonal phase transition in the prototypical MHP compound, MAPbI$_3$. The observation is largely enabled by the charge-excitation-free nature of the experiment, which yields clean signature of the phase transition. The tetragonal phase is found to develop in the tens of ns timescale before returning to the orthorhombic phase over the course of a few hundred ns — a pace dictated by heat conduction from the film to the substrate. Comparison between MAPbI$_3$ films with different steady-state $T_{o-t}$ reveals that films that phase transition at a lower temperature exhibit a faster transient *o*-to-*t* phase transition. When corroborated by first principles calculations such as ab-initio and/or machine-learning enabled molecular dynamics simulations,[21, 61] the reported approach here to experimentally probe the fundamental timescales of phase transition in MHPs may shed light onto the energetic barriers of phase transitions in MHPs. Investigation of photothermally induced phase transitions in MAPbI$_3$ or other hybrid MHPs may further enable the design of optically triggered, large-amplitude dielectric switches, as it has been reported that the dielectric permittivity of MAPbX$_3$ (X=Cl, Br or I) increases by as much as 3~5 times when going from the orthorhombic to the tetragonal phase.[62] The impulsive vibrational pumping approach can also be generalized to the investigation of phase transitions in other hybrid and organic materials, such as spin crossover materials[63] as well as superatomic crystals,[64] for potential applications towards spintronics, nonvolatile memory, and thermal switching. Compared to other techniques to manipulate transient phase transitions such as the ultrafast-heating calorimetry and μs-electric pulsed Joule-heating,[65, 66] which have heating rates in the $10^4$~$10^7$ K·s$^{-1}$ range,[67] fs vibrational pumping provides a much higher heating rate in the $10^9 \sim 10^{10}$ K·s$^{-1}$ (1~10 K temperature rise in ~1 ns) and therefore a higher temporal resolution for revealing the fundamental timescales of phase transitions in solid-state hybrid and organic materials.

**Methods**

*Sample fabrication.* The fabrication of MAPbI$_3$ film on CaF$_2$ substrate followed previous work.[68] The MAPbI$_3$ precursor solution was prepared by dissolving 159 mg of methylammonium iodide (MAI; Greatcell Solar, Australia), 461 mg of Pb iodide (PbI$_2$; 99.99%, TCI) and 78 mg of dimethyl sulfoxide (DMSO; 99.7%, Acros Organics, USA) in 500 mg of *N,N*-dimethylformamide (DMF; 99.8%, Acros Organics, USA). The substrates were treated with UV-ozone for 15 min to ensure good solution wettability. The MAPbI$_3$ layer was deposited by spin-coating the as-prepared precursor solution at 4000 rpm for 30s with an acceleration of 1300 rpm·s$^{-1}$. At the 10$^{th}$ second of spinning, 250 μL of anti-solvent diethyl ether (DE; 99.7%, Sigma-Aldrich) were dripped at the center. Subsequently, the as-deposited films were annealed at 100 °C for 20 min. This fabrication method consistently yields MAPbI$_3$ films with thickness of ~600 nm. The fabrication of MAPbI$_3$ films on sapphire substrate was based on a different literature report.[69] The 40 wt% MAPbI$_3$ precursor solution was prepared by dissolving 1.2:1 molar ratio of methylammonium iodide (MAI; Greatcell Solar, Australia), and lead iodide (PbI$_2$; 99.99%, TCI) in a mixed solvent of N-Methyl-2-pyrrolidone (NMP; 99.5%, Sigma-Aldrich, USA) and γ-Butyrolactone (GBL; 99%, Sigma-Aldrich, USA) (volumetric ratio 1:1). The MAPbI$_3$ layer was deposited by spin-coating the as-prepared precursor solution at 3000 rpm for 30 s. Subsequently, the as-deposited films were rapidly immersed in a diethyl ether (DE; 99%, Sigma-Aldrich, USA) bath for 90 s, then annealed at 150 °C for 15 mins in a N$_2$-filled glovebox.

*Static characterization.* A high-resolution scanning electron microscope (SEM; Quattro ESEM, ThermoFisher Scientific, USA) was used to obtain cross-sectional images of the samples as well as to quantify the film thickness. For steady-state transmittance measurements in the visible range, a deuterium-halogen light source (AvaLight-DHc-S, Avantes) and a fiber-coupled spectrometer (AvaSpec-ULS2048CL-EVO, Avantes) were used. The steady-state infrared absorbance were measured using FTIR (Nicolet 6700).

*Transient absorption experiments*. The mid-IR pump pulses were produced by a high-energy mid-IR optical parametric amplifier (Orpheus-One-HE), which is pumped by a Pharos amplifier (170 fs pulse duration and 1030 nm pulse wavelength) with 0.9 mJ pulse energy at 1.5 kHz repetition rate. The visible pump pulses at 700 nm were generated by a separate optical parametric amplifier (Orpheus-F) pumped by the same Pharos amplifier with 0.5 mJ input pulse energy. The broadband visible probe pulses were generated by a supercontinuum ns laser (NKT Compact) and triggered and electrically delayed by a digital delay generator (SRS DG645). The instrumental timing jitter of the probe pulses was eliminated by an event timer (Picoquant MultiHarp 150P), which measures the actual arrival times of all the pump and the probe pulses during the transient measurements. Cryogenic measurements were enabled by a liquid-nitrogen optical cryostat (Janis VPF-100) at a vacuum level better than $10^{-4}$ Torr.

**Supporting Information**

Additional figures and discussion can be found in the Supporting Information.

**Correspondence**

E-mail: peijun.guo@yale.edu

**Acknowledgements**

Work at Yale is supported by the Yale University lab set-up fund. Z.D. and N.P.P. acknowledge the support from the US Office of Naval Research (grant no. N00014-20-1-2574). We thank Prof. Amir Haji-Akbari from Yale University and Prof. Yao Wang from Clemson University for insightful discussions.

**Data Availability**

The data that support the findings of this study are available from the corresponding author upon reasonable request.

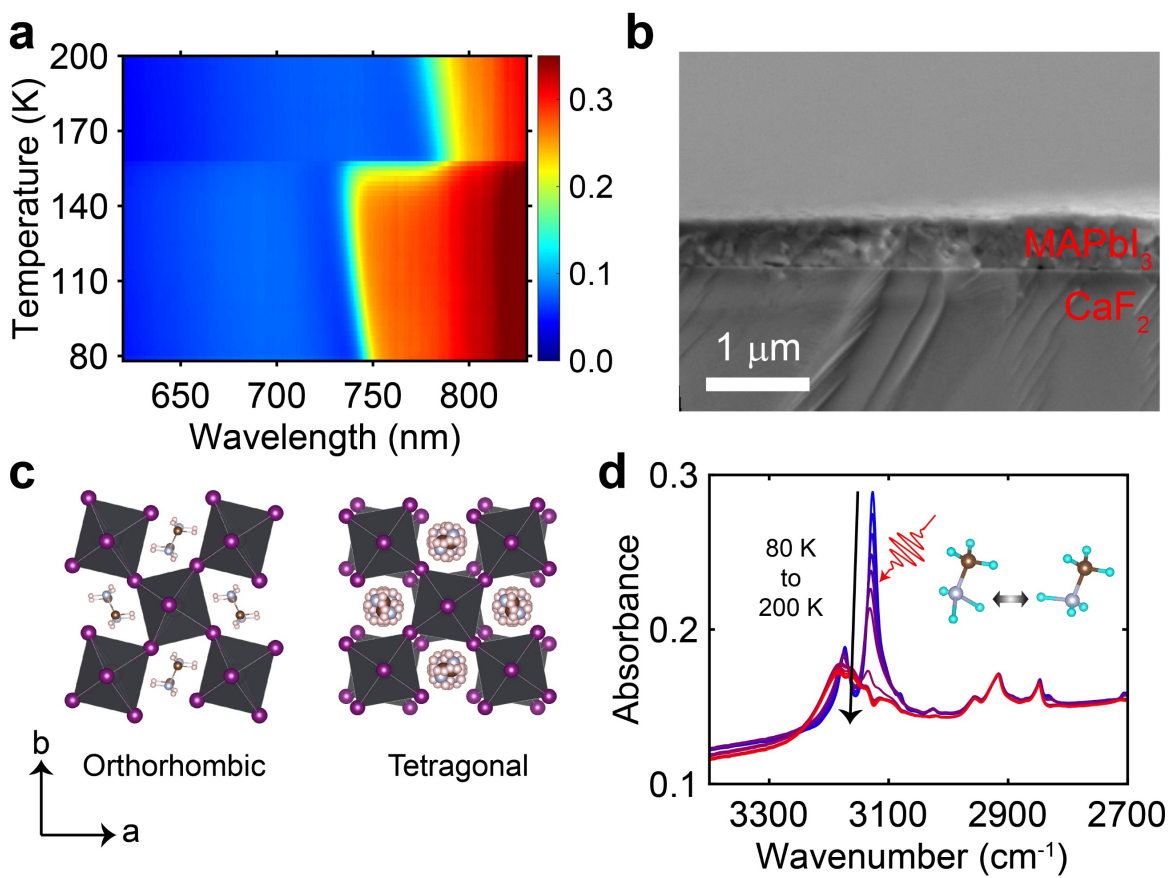

**Fig. 1**. **a** Steady-state optical transmittance spectra as a function of temperature for an MAPbI$_3$ film on a CaF$_2$ substrate. **b** Cross-sectional scanning electron microscopy image of the ~600 nm thick MAPbI$_3$ film on CaF$_2$ substrate. **c** Crystal structures of the orthorhombic and tetragonal phases of MAPbI$_3$ (carbon: brown; nitrogen: gray); note that two layers of cations along *c* are presented in the plots, which are ordered in the orthorhombic phase but dynamically disordered in the tetragonal phase. **d** Steady-state infrared absorbance of the MAPbI$_3$ film on the CaF$_2$ substrate (plotted in terms of the optical density).

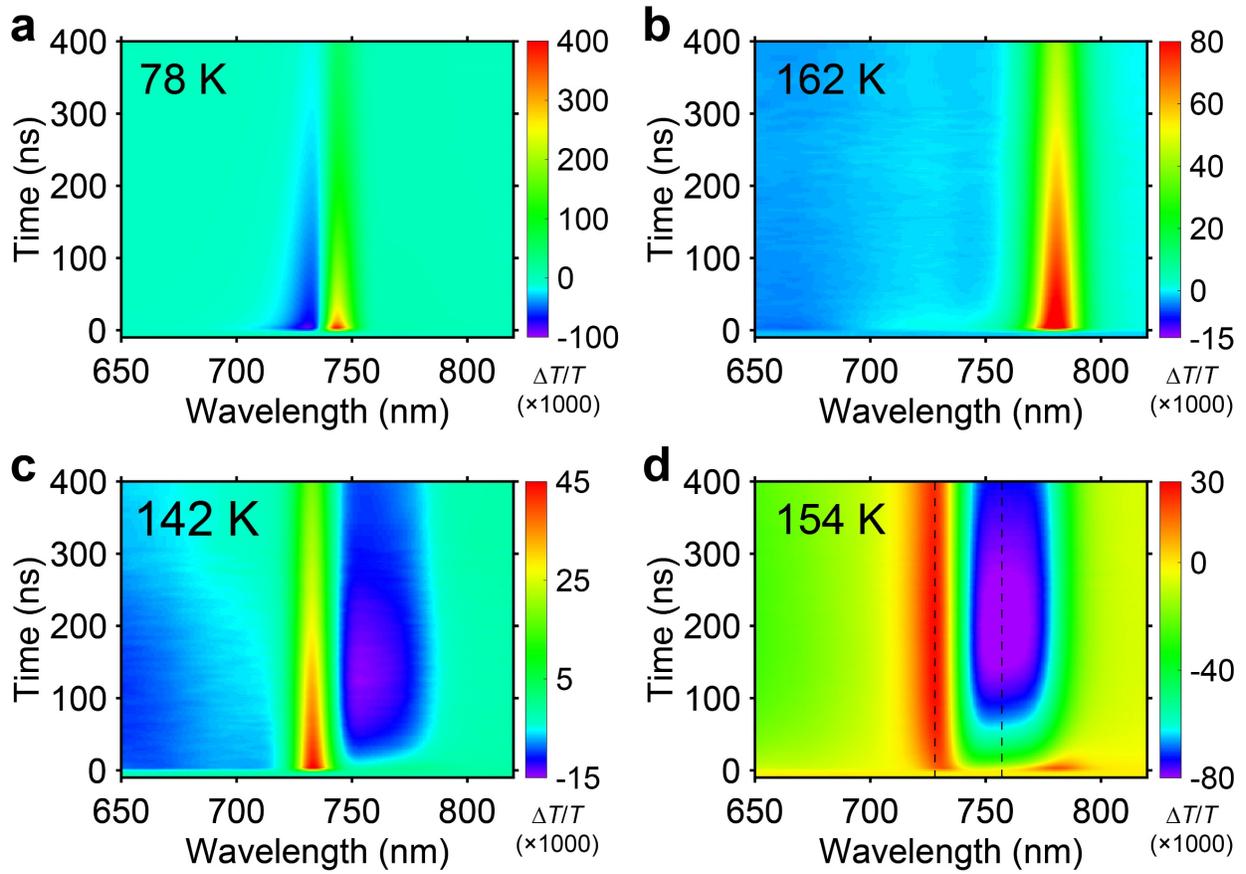

**Fig. 2**. Transient $\Delta T/T$ spectral map measured for the MAPbI$_3$ film on CaF$_2$ at temperature of 78 K (orthorhombic phase) in **a**, 162 K (tetragonal phase) in **b**, 142 K in **c** and 154 K in **d**. The pump wavelength was 3170 nm and the fluence was 2.4 mJ·cm$^{-2}$.

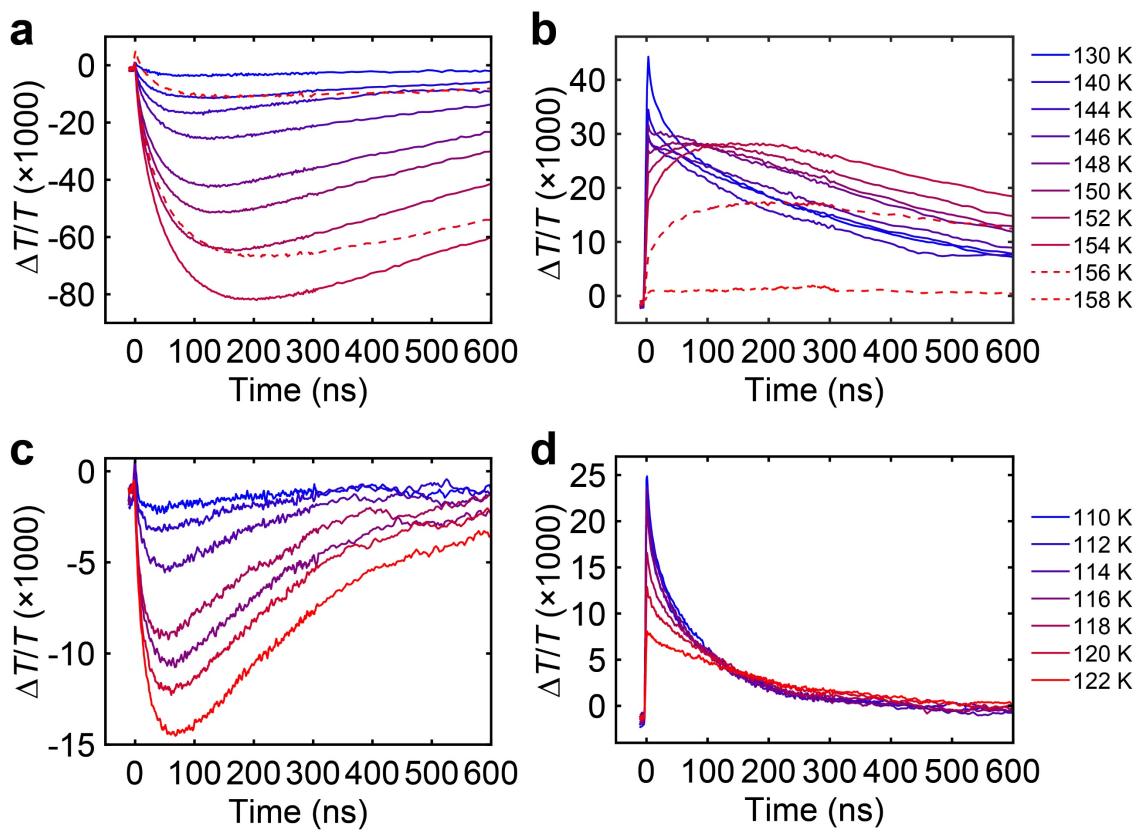

**Fig. 3.** Transient Δ*T*/*T* kinetics extracted at 757 nm (in **a**) and at 728 nm (in **b**) for MAPbI$_3$ on CaF$_2$ as a function of sample temperature. Transient Δ*T*/*T* kinetics extracted at 757 nm (in **c**) and at 728 nm (in **d**) for MAPbI$_3$ on sapphire as a function of sample temperature.

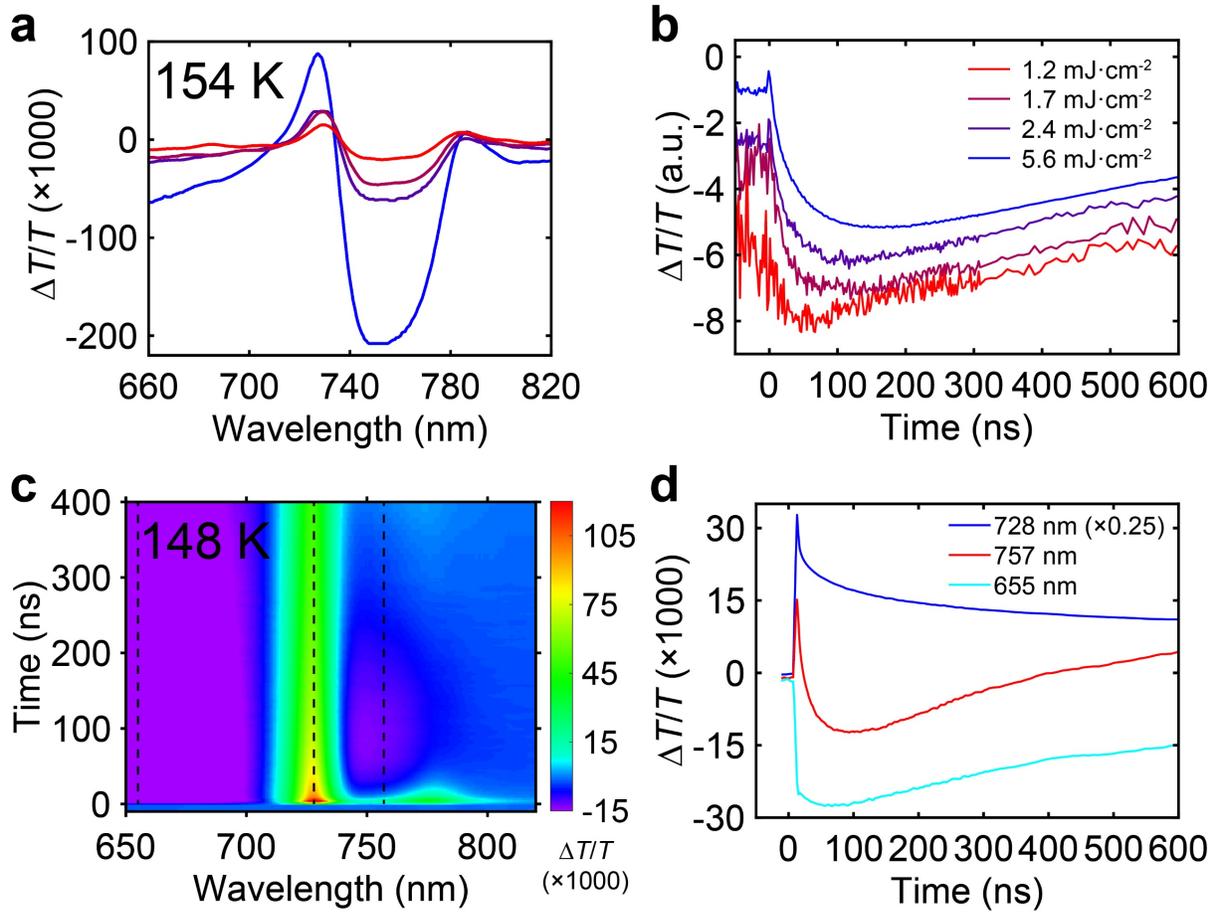

**Fig. 4**. Fluence dependent $\Delta T/T$ spectra at 200 ns delay time in **a** and fluence dependent $\Delta T/T$ kinetics at 757 nm in **b**. The results are measured for MAPbI$_3$ on CaF$_2$ at 154 K. The legend in **b** also applies to **a**. **c** Transient $\Delta T/T$ spectral map of for MAPbI$_3$ on CaF$_2$ measured using a 700-nm pump pulses at 148 K, using a pump fluence of 2.5 mJ·cm$^{-2}$. **d** $\Delta T/T$ kinetics plotted at three representative wavelengths (728 nm, 757 nm and 655 nm, as indicated by the black dashed lines in **c**).

**References**

1. Zhou, H.; Chen, Q.; Li, G.; Luo, S.; Song, T.-b.; Duan, H.-S.; Hong, Z.; You, J.; Liu, Y.; Yang, Y., Interface engineering of highly efficient perovskite solar cells. *Science* **2014,** *345* (6196), 542-546.

2. Yuan, M.; Quan, L. N.; Comin, R.; Walters, G.; Sabatini, R.; Voznyy, O.; Hoogland, S.; Zhao, Y.; Beauregard, E. M.; Kanjanaboos, P.; Lu, Z.; Kim, D. H.; Sargent, E. H., Perovskite energy funnels for efficient light-emitting diodes. *Nature Nanotechnology* **2016,** *11* (10), 872-877.

3. Kim, Y.-H.; Zhai, Y.; Lu, H.; Pan, X.; Xiao, C.; Gaulding, E. A.; Harvey, S. P.; Berry, J. J.; Vardeny, Z. V.; Luther, J. M.; Beard, M. C., Chiral-induced spin selectivity enables a room-temperature spin light-emitting diode. *Science* **2021,** *371* (6534), 1129-1133.

4. Lee, M. M.; Teuscher, J.; Miyasaka, T.; Murakami, T. N.; Snaith, H. J., Efficient Hybrid Solar Cells Based on Meso-Superstructured Organometal Halide Perovskites. *Science* **2012,** *338* (6107), 643-647.

5. Park, N.-G.; Grätzel, M.; Miyasaka, T.; Zhu, K.; Emery, K., Towards stable and commercially available perovskite solar cells. *Nature Energy* **2016,** *1* (11), 16152.

6. Tsai, H.; Nie, W.; Blancon, J.-C.; Stoumpos, C. C.; Asadpour, R.; Harutyunyan, B.; Neukirch, A. J.; Verduzco, R.; Crochet, J. J.; Tretiak, S.; Pedesseau, L.; Even, J.; Alam, M. A.; Gupta, G.; Lou, J.; Ajayan, P. M.; Bedzyk, M. J.; Kanatzidis, M. G.; Mohite, A. D., High-efficiency two-dimensional Ruddlesden–Popper perovskite solar cells. *Nature* **2016,** *536* (7616), 312-316.

7. Sun, B.; Liu, X.-F.; Li, X.-Y.; Zhang, Y.; Shao, X.; Yang, D.; Zhang, H.-L., Two-Dimensional Perovskite Chiral Ferromagnets. *Chemistry of Materials* **2020,** *32* (20), 8914-8920.

8. Shi, E.; Yuan, B.; Shiring, S. B.; Gao, Y.; Akriti; Guo, Y.; Su, C.; Lai, M.; Yang, P.; Kong, J.; Savoie, B. M.; Yu, Y.; Dou, L., Two-dimensional halide perovskite lateral epitaxial heterostructures. *Nature* **2020,** *580* (7805), 614-620.

9. Stoumpos, C. C.; Malliakas, C. D.; Kanatzidis, M. G., Semiconducting Tin and Lead Iodide Perovskites with Organic Cations: Phase Transitions, High Mobilities, and Near-Infrared Photoluminescent Properties. *Inorganic Chemistry* **2013,** *52* (15), 9019-9038.

10. Boyd, C. C.; Cheacharoen, R.; Leijtens, T.; McGehee, M. D., Understanding Degradation Mechanisms and Improving Stability of Perovskite Photovoltaics. *Chemical Reviews* **2019,** *119* (5), 3418-3451.

11. Lin, J.; Lai, M.; Dou, L.; Kley, C. S.; Chen, H.; Peng, F.; Sun, J.; Lu, D.; Hawks, S. A.; Xie, C.; Cui, F.; Alivisatos, A. P.; Limmer, D. T.; Yang, P., Thermochromic halide perovskite solar cells. *Nature Materials* **2018,** *17* (3), 261-267.

12. Zhumekenov, A. A.; Saidaminov, M. I.; Mohammed, O. F.; Bakr, O. M., Stimuli-responsive switchable halide perovskites: Taking advantage of instability. *Joule* **2021,** *5* (8), 2027-2046.

13. Xu, L.-J.; Lin, H.; Lee, S.; Zhou, C.; Worku, M.; Chaaban, M.; He, Q.; Plaviak, A.; Lin, X.; Chen, B.; Du, M.-H.; Ma, B., 0D and 2D: The Cases of Phenylethylammonium Tin Bromide Hybrids. *Chemistry of Materials* **2020,** *32* (11), 4692-4698.

14. Straus, D. B.; Guo, S.; Cava, R. J., Kinetically Stable Single Crystals of Perovskite-Phase $CsPbI_3$. *Journal of the American Chemical Society* **2019,** *141* (29), 11435-11439.

15. Smith, I. C.; Smith, M. D.; Jaffe, A.; Lin, Y.; Karunadasa, H. I., Between the Sheets: Postsynthetic Transformations in Hybrid Perovskites. *Chemistry of Materials* **2017,** *29* (5), 1868-1884.

16. Weller, M. T.; Weber, O. J.; Henry, P. F.; Di Pumpo, A. M.; Hansen, T. C., Complete structure and cation orientation in the perovskite photovoltaic methylammonium lead iodide between 100 and 352 K. *Chemical Communications* **2015,** *51* (20), 4180-4183.

17. Whitfield, P. S.; Herron, N.; Guise, W. E.; Page, K.; Cheng, Y. Q.; Milas, I.; Crawford, M. K., Structures, Phase Transitions and Tricritical Behavior of the Hybrid Perovskite Methyl Ammonium Lead Iodide. *Scientific Reports* **2016,** *6* (1), 35685.

18. Yaffe, O.; Guo, Y.; Tan, L. Z.; Egger, D. A.; Hull, T.; Stoumpos, C. C.; Zheng, F.; Heinz, T. F.; Kronik, L.; Kanatzidis, M. G.; Owen, J. S.; Rappe, A. M.; Pimenta, M. A.; Brus, L. E., Local Polar Fluctuations in Lead Halide Perovskite Crystals. *Physical Review Letters* **2017,** *118* (13), 136001.

19. Onoda-Yamamuro, N.; Matsuo, T.; Suga, H., Calorimetric and IR spectroscopic studies of phase transitions in methylammonium trihalogenoplumbates (II)†. *Journal of Physics and Chemistry of Solids* **1990,** *51* (12), 1383-1395.

20. Lai, M.; Lei, T.; Zhang, Y.; Jin, J.; Steele, J. A.; Yang, P., Phase transition dynamics in one-dimensional halide perovskite crystals. *MRS Bulletin* **2021,** *46* (4), 310-316.

21. Bischak, C. G.; Lai, M.; Fan, Z.; Lu, D.; David, P.; Dong, D.; Chen, H.; Etman, A. S.; Lei, T.; Sun, J.; Grünwald, M.; Limmer, D. T.; Yang, P.; Ginsberg, N. S., Liquid-like Interfaces Mediate Structural Phase Transitions in Lead Halide Perovskites. *Matter* **2020,** *3* (2), 534-545.

22. Lin, Z.; Zhang, Y.; Gao, M.; Steele, J. A.; Louisia, S.; Yu, S.; Quan, L. N.; Lin, C.-K.; Limmer, D. T.; Yang, P., Kinetics of moisture-induced phase transformation in inorganic halide perovskite. *Matter* **2021,** *4* (7), 2392-2402.

23. Cavalleri, A.; Tóth, C.; Siders, C. W.; Squier, J. A.; Ráksi, F.; Forget, P.; Kieffer, J. C., Femtosecond Structural Dynamics in $VO_2$ during an Ultrafast Solid-Solid Phase Transition. *Physical Review Letters* **2001,** *87* (23), 237401.

24. Frohna, K.; Deshpande, T.; Harter, J.; Peng, W.; Barker, B. A.; Neaton, J. B.; Louie, S. G.; Bakr, O. M.; Hsieh, D.; Bernardi, M., Inversion symmetry and bulk Rashba effect in methylammonium lead iodide perovskite single crystals. *Nature Communications* **2018,** *9* (1), 1829.

25. Davies, C. L.; Filip, M. R.; Patel, J. B.; Crothers, T. W.; Verdi, C.; Wright, A. D.; Milot, R. L.; Giustino, F.; Johnston, M. B.; Herz, L. M., Bimolecular recombination in methylammonium lead triiodide perovskite is an inverse absorption process. *Nature Communications* **2018,** *9* (1), 293.

26. Quarti, C.; Mosconi, E.; Ball, J. M.; D'Innocenzo, V.; Tao, C.; Pathak, S.; Snaith, H. J.; Petrozza, A.; De Angelis, F., Structural and optical properties of methylammonium lead iodide across the tetragonal to cubic phase transition: implications for perovskite solar cells. *Energy & Environmental Science* **2016,** *9* (1), 155-163.

27. Leguy, A. M. A.; Frost, J. M.; McMahon, A. P.; Sakai, V. G.; Kockelmann, W.; Law, C.; Li, X.; Foglia, F.; Walsh, A.; O'Regan, B. C.; Nelson, J.; Cabral, J. T.; Barnes, P. R. F., The dynamics of methylammonium ions in hybrid organic–inorganic perovskite solar cells. *Nature Communications* **2015,** *6* (1), 7124.

28. Poglitsch, A.; Weber, D., Dynamic disorder in methylammoniumtrihalogenoplumbates (II) observed by millimeter‐wave spectroscopy. *The Journal of Chemical Physics* **1987,** *87* (11), 6373-6378.

29. Brivio, F.; Frost, J. M.; Skelton, J. M.; Jackson, A. J.; Weber, O. J.; Weller, M. T.; Goñi, A. R.; Leguy, A. M. A.; Barnes, P. R. F.; Walsh, A., Lattice dynamics and vibrational spectra of the orthorhombic, tetragonal, and cubic phases of methylammonium lead iodide. *Physical Review B* **2015,** *92* (14), 144308.

30. Guo, P.; Mannodi-Kanakkithodi, A.; Gong, J.; Xia, Y.; Stoumpos, C. C.; Cao, D. H.; Diroll, B. T.; Ketterson, J. B.; Wiederrecht, G. P.; Xu, T.; Chan, M. K. Y.; Kanatzidis, M. G.; Schaller,

R. D., Infrared-pump electronic-probe of methylammonium lead iodide reveals electronically decoupled organic and inorganic sublattices. *Nature Communications* **2019,** *10* (1), 482.

31. Zhou, Y.; Poli, I.; Meggiolaro, D.; De Angelis, F.; Petrozza, A., Defect activity in metal halide perovskites with wide and narrow bandgap. *Nature Reviews Materials* **2021,** *6* (11), 986-1002.

32. Dobrovolsky, A.; Merdasa, A.; Unger, E. L.; Yartsev, A.; Scheblykin, I. G., Defect-induced local variation of crystal phase transition temperature in metal-halide perovskites. *Nature Communications* **2017,** *8* (1), 34.

33. Strelcov, E.; Dong, Q.; Li, T.; Chae, J.; Shao, Y.; Deng, Y.; Gruverman, A.; Huang, J.; Centrone, A., $CH_3NH_3PbI_3$ perovskites: Ferroelasticity revealed. *Science Advances* **2017,** *3* (4), e1602165.

34. Wen, H.; Cherukara, M. J.; Holt, M. V., Time-Resolved X-Ray Microscopy for Materials Science. *Annual Review of Materials Research* **2019,** *49* (1), 389-415.

35. Pisoni, A.; Jaćimović, J.; Barišić, O. S.; Spina, M.; Gaál, R.; Forró, L.; Horváth, E., Ultra-Low Thermal Conductivity in Organic–Inorganic Hybrid Perovskite $CH_3NH_3PbI_3$. *The Journal of Physical Chemistry Letters* **2014,** *5* (14), 2488-2492.

36. Li, D.; Wang, G.; Cheng, H.-C.; Chen, C.-Y.; Wu, H.; Liu, Y.; Huang, Y.; Duan, X., Size-dependent phase transition in methylammonium lead iodide perovskite microplate crystals. *Nature Communications* **2016,** *7* (1), 11330.

37. Stavrakas, C.; Zelewski, S. J.; Frohna, K.; Booker, E. P.; Galkowski, K.; Ji, K.; Ruggeri, E.; Mackowski, S.; Kudrawiec, R.; Plochocka, P.; Stranks, S. D., Influence of Grain Size on Phase Transitions in Halide Perovskite Films. *Advanced Energy Materials* **2019,** *9* (35), 1901883.

38. Schötz, K.; Askar, A. M.; Köhler, A.; Shankar, K.; Panzer, F., Investigating the Tetragonal-to-Orthorhombic Phase Transition of Methylammonium Lead Iodide Single Crystals by Detailed Photoluminescence Analysis. *Advanced Optical Materials* **2020,** *8* (17), 2000455.

39. Wright, A. D.; Verdi, C.; Milot, R. L.; Eperon, G. E.; Pérez-Osorio, M. A.; Snaith, H. J.; Giustino, F.; Johnston, M. B.; Herz, L. M., Electron–phonon coupling in hybrid lead halide perovskites. *Nature Communications* **2016,** *7* (1), 11755.

40. Straus, D. B.; Hurtado Parra, S.; Iotov, N.; Gebhardt, J.; Rappe, A. M.; Subotnik, J. E.; Kikkawa, J. M.; Kagan, C. R., Direct Observation of Electron–Phonon Coupling and Slow Vibrational Relaxation in Organic–Inorganic Hybrid Perovskites. *Journal of the American Chemical Society* **2016,** *138* (42), 13798-13801.

41. Wu, X.; Trinh, M. T.; Niesner, D.; Zhu, H.; Norman, Z.; Owen, J. S.; Yaffe, O.; Kudisch, B. J.; Zhu, X. Y., Trap States in Lead Iodide Perovskites. *Journal of the American Chemical Society* **2015,** *137* (5), 2089-2096.

42. Yang, Y.; Yang, M.; Li, Z.; Crisp, R.; Zhu, K.; Beard, M. C., Comparison of Recombination Dynamics in $CH_3NH_3PbBr_3$ and $CH_3NH_3PbI_3$ Perovskite Films: Influence of Exciton Binding Energy. *The Journal of Physical Chemistry Letters* **2015,** *6* (23), 4688-4692.

43. Guo, Z.; Wu, X.; Zhu, T.; Zhu, X.; Huang, L., Electron–Phonon Scattering in Atomically Thin 2D Perovskites. *ACS Nano* **2016,** *10* (11), 9992-9998.

44. Price, M. B.; Butkus, J.; Jellicoe, T. C.; Sadhanala, A.; Briane, A.; Halpert, J. E.; Broch, K.; Hodgkiss, J. M.; Friend, R. H.; Deschler, F., Hot-carrier cooling and photoinduced refractive index changes in organic–inorganic lead halide perovskites. *Nature Communications* **2015,** *6* (1), 8420.

45. Yang, Z.; Surrente, A.; Galkowski, K.; Bruyant, N.; Maude, D. K.; Haghighirad, A. A.; Snaith, H. J.; Plochocka, P.; Nicholas, R. J., Unraveling the Exciton Binding Energy and the Dielectric Constant in Single-Crystal Methylammonium Lead Triiodide Perovskite. *The Journal of Physical Chemistry Letters* **2017,** *8* (8), 1851-1855.

46. Guzelturk, B.; Winkler, T.; Van de Goor, T. W. J.; Smith, M. D.; Bourelle, S. A.; Feldmann, S.; Trigo, M.; Teitelbaum, S. W.; Steinrück, H.-G.; de la Pena, G. A.; Alonso-Mori, R.; Zhu, D.; Sato, T.; Karunadasa, H. I.; Toney, M. F.; Deschler, F.; Lindenberg, A. M., Visualization of dynamic polaronic strain fields in hybrid lead halide perovskites. *Nature Materials* **2021,** *20* (5), 618-623.

47. Jia, Y.; Kerner, R. A.; Grede, A. J.; Rand, B. P.; Giebink, N. C., Continuous-wave lasing in an organic–inorganic lead halide perovskite semiconductor. *Nature Photonics* **2017,** *11* (12), 784-788.

48. Huang, Y.; Yang, S.; Teitelbaum, S.; De la Peña, G.; Sato, T.; Chollet, M.; Zhu, D.; Niedziela, J. L.; Bansal, D.; May, A. F.; Lindenberg, A. M.; Delaire, O.; Reis, D. A.; Trigo, M., Observation of a Novel Lattice Instability in Ultrafast Photoexcited SnSe. *Physical Review X* **2022,** *12* (1), 011029.

49. Zong, A.; Kogar, A.; Bie, Y.-Q.; Rohwer, T.; Lee, C.; Baldini, E.; Ergeçen, E.; Yilmaz, M. B.; Freelon, B.; Sie, E. J.; Zhou, H.; Straquadine, J.; Walmsley, P.; Dolgirev, P. E.; Rozhkov, A. V.; Fisher, I. R.; Jarillo-Herrero, P.; Fine, B. V.; Gedik, N., Evidence for topological defects in a photoinduced phase transition. *Nature Physics* **2019,** *15* (1), 27-31.

50. Zhang, J.; Tan, X.; Liu, M.; Teitelbaum, S. W.; Post, K. W.; Jin, F.; Nelson, K. A.; Basov, D. N.; Wu, W.; Averitt, R. D., Cooperative photoinduced metastable phase control in strained manganite films. *Nature Materials* **2016,** *15* (9), 956-960.

51. Guo, P.; Weimer, M. S.; Emery, J. D.; Diroll, B. T.; Chen, X.; Hock, A. S.; Chang, R. P. H.; Martinson, A. B. F.; Schaller, R. D., Conformal Coating of a Phase Change Material on Ordered Plasmonic Nanorod Arrays for Broadband All-Optical Switching. *ACS Nano* **2017,** *11* (1), 693-701.

52. Beaud, P.; Caviezel, A.; Mariager, S. O.; Rettig, L.; Ingold, G.; Dornes, C.; Huang, S. W.; Johnson, J. A.; Radovic, M.; Huber, T.; Kubacka, T.; Ferrer, A.; Lemke, H. T.; Chollet, M.; Zhu, D.; Glownia, J. M.; Sikorski, M.; Robert, A.; Wadati, H.; Nakamura, M.; Kawasaki, M.; Tokura, Y.; Johnson, S. L.; Staub, U., A time-dependent order parameter for ultrafast photoinduced phase transitions. *Nature Materials* **2014,** *13* (10), 923-927.

53. Li, X.; Qiu, T.; Zhang, J.; Baldini, E.; Lu, J.; Rappe, A. M.; Nelson, K. A., Terahertz field-induced ferroelectricity in quantum paraelectric $SrTiO_3$. *Science* **2019,** *364* (6445), 1079-1082.

54. Saha, T.; Golež, D.; De Ninno, G.; Mravlje, J.; Murakami, Y.; Ressel, B.; Stupar, M.; Ribič, P. R., Photoinduced phase transition and associated timescales in the excitonic insulator $Ta_2NiSe_5$. *Physical Review B* **2021,** *103* (14), 144304.

55. Zong, A.; Dolgirev, P. E.; Kogar, A.; Ergeçen, E.; Yilmaz, M. B.; Bie, Y.-Q.; Rohwer, T.; Tung, I. C.; Straquadine, J.; Wang, X.; Yang, Y.; Shen, X.; Li, R.; Yang, J.; Park, S.; Hoffmann, M. C.; Ofori-Okai, B. K.; Kozina, M. E.; Wen, H.; Wang, X.; Fisher, I. R.; Jarillo-Herrero, P.; Gedik, N., Dynamical Slowing-Down in an Ultrafast Photoinduced Phase Transition. *Physical Review Letters* **2019,** *123* (9), 097601.

56. Kirschner, M. S.; Diroll, B. T.; Guo, P.; Harvey, S. M.; Helweh, W.; Flanders, N. C.; Brumberg, A.; Watkins, N. E.; Leonard, A. A.; Evans, A. M.; Wasielewski, M. R.; Dichtel, W. R.; Zhang, X.; Chen, L. X.; Schaller, R. D., Photoinduced, reversible phase transitions in all-inorganic perovskite nanocrystals. *Nature Communications* **2019,** *10* (1), 504.

57. Sharma, R.; Dai, Z.; Gao, L.; Brenner, T. M.; Yadgarov, L.; Zhang, J.; Rakita, Y.; Korobko, R.; Rappe, A. M.; Yaffe, O., Elucidating the atomistic origin of anharmonicity in tetragonal $CH_3NH_3PbI_3$ with Raman scattering. *Physical Review Materials* **2020,** *4* (9), 092401.

58. Chen, T.; Foley, B. J.; Park, C.; Brown, C. M.; Harriger, L. W.; Lee, J.; Ruff, J.; Yoon, M.; Choi, J. J.; Lee, S.-H., Entropy-driven structural transition and kinetic trapping in formamidinium lead iodide perovskite. *Science Advances* **2016,** *2* (10), e1601650.

59. Sutton, R. J.; Filip, M. R.; Haghighirad, A. A.; Sakai, N.; Wenger, B.; Giustino, F.; Snaith, H. J., Cubic or Orthorhombic? Revealing the Crystal Structure of Metastable Black-Phase $CsPbI_3$ by Theory and Experiment. *ACS Energy Letters* **2018,** *3* (8), 1787-1794.

60. Ke, F.; Wang, C.; Jia, C.; Wolf, N. R.; Yan, J.; Niu, S.; Devereaux, T. P.; Karunadasa, H. I.; Mao, W. L.; Lin, Y., Preserving a robust $CsPbI_3$ perovskite phase via pressure-directed octahedral tilt. *Nature Communications* **2021,** *12* (1), 461.

61. Jinnouchi, R.; Lahnsteiner, J.; Karsai, F.; Kresse, G.; Bokdam, M., Phase Transitions of Hybrid Perovskites Simulated by Machine-Learning Force Fields Trained on the Fly with Bayesian Inference. *Physical Review Letters* **2019,** *122* (22), 225701.

62. Onoda-Yamamuro, N.; Matsuo, T.; Suga, H., Dielectric study of $CH_3NH_3PbX_3$ (X = Cl, Br, I). *Journal of Physics and Chemistry of Solids* **1992,** *53* (7), 935-939.

63. Gaspar, A. B.; Ksenofontov, V.; Seredyuk, M.; Gütlich, P., Multifunctionality in spin crossover materials. *Coordination Chemistry Reviews* **2005,** *249* (23), 2661-2676.

64. Ong, W.-L.; O'Brien, E. S.; Dougherty, P. S. M.; Paley, D. W.; Fred Higgs Iii, C.; McGaughey, A. J. H.; Malen, J. A.; Roy, X., Orientational order controls crystalline and amorphous thermal transport in superatomic crystals. *Nature Materials* **2017,** *16* (1), 83-88.

65. Dana, A.; Sekiguchi, H.; Aoyama, K.; Faran, E.; Liss, K.-D.; Shilo, D., The evolution of the martensitic transformation at the high-driving-force regime: A microsecond-scale time-resolved X-ray diffraction study. *Journal of Alloys and Compounds* **2021,** *856*, 157968.

66. Zhang, Y.; Fowler, C.; Liang, J.; Azhar, B.; Shalaginov, M. Y.; Deckoff-Jones, S.; An, S.; Chou, J. B.; Roberts, C. M.; Liberman, V.; Kang, M.; Ríos, C.; Richardson, K. A.; Rivero-Baleine, C.; Gu, T.; Zhang, H.; Hu, J., Electrically reconfigurable non-volatile metasurface using low-loss optical phase-change material. *Nature Nanotechnology* **2021,** *16* (6), 661-666.

67. Orava, J.; Greer, A. L.; Gholipour, B.; Hewak, D. W.; Smith, C. E., Characterization of supercooled liquid $Ge_2Sb_2Te_5$ and its crystallization by ultrafast-heating calorimetry. *Nature Materials* **2012,** *11* (4), 279-283.

68. Ahn, N.; Son, D.-Y.; Jang, I.-H.; Kang, S. M.; Choi, M.; Park, N.-G., Highly Reproducible Perovskite Solar Cells with Average Efficiency of 18.3% and Best Efficiency of 19.7% Fabricated via Lewis Base Adduct of Lead(II) Iodide. *Journal of the American Chemical Society* **2015,** *137* (27), 8696-8699.

69. Yang, M.; Zhou, Y.; Zeng, Y.; Jiang, C.-S.; Padture, N. P.; Zhu, K., Square-Centimeter Solution-Processed Planar $CH_3NH_3PbI_3$ Perovskite Solar Cells with Efficiency Exceeding 15%. *Advanced Materials* **2015,** *27* (41), 6363-6370.